\documentclass[]{elsart}
\usepackage{epsfig,latexsym}

\newcommand{\sm}[1]{\mbox{{\scriptsize #1}}}
\newcommand{\simle} {\,{}^<_{\sim}\,}
\newcommand{\simge} {\,{}^>_{\sim}\,}

\begin{document}
\begin{frontmatter}

\title{B-ball Dark Matter and Baryogenesis}
\author{Robi Banerjee
\and Karsten Jedamzik\thanksref{email}}
\address{Max--Planck--Institut
f\"ur Astrophysik 
Karl--Schwarzschild--Str. 1
85741 Garching, Germany}
\thanks[email]{e-mail: banerjee@mpa-garching.mpg.de, 
jedamzik@mpa-garching.mpg.de}

\begin{abstract}
It has been recently suggested that stable, supersymmetric
B-balls formed in the early universe could not only be the dark matter
at the present epoch, but also be responsible for baryogenesis by
their partial evaporation at high temperatures. We reinvestigate the
efficiency of B-ball baryogenesis and find it to be limited by the diffusion
of baryon number away from the B-balls. Successful baryogenesis may only
occur for B-balls with charges $Q\simle 10^{20} - 5\times 10^{23}$,
which is close to the observational lower limits on the $Q$ of a
significant
B-ball dark matter component. We also present some cosmological
constraints on the abundances of larger B-balls in the early
universe.
\\

\end{abstract}
\end{frontmatter}

It is well known that particle physics models containing 
an unbroken U(1) symmetry allow for the existence of
non-topological solitons \cite{fls,col}, i.e. Q-balls, which carry a large
number of a conserved global charge \cite{dks}. If the effective
potential $U(\Phi)$, of the scalar $\Phi$ carrying the global charge 
grows slower than the second power of $\Phi$, the
mass of the solitonic object scales with the U(1) charge $Q$ as
$M_Q \approx \tilde{M}\,Q^p$ ($0<p<1$), where $\tilde{M}$ 
is some energy scale.   
The minimal supersymmetric standard model (MSSM) with
supersymmetry breaking communicated at low energy scale
contains an effective potential which
is nearly flat $\sim M^4$ at large $\Phi$. In this case, 
$M_Q \approx M\,Q^{3/4}$, where $M\approx 1 - 10$ TeV is the
SUSY breaking scale \cite{dks}. Such Q-balls are absolutely
stable at zero temperature if their mass $M_Q$ becomes smaller 
than the total mass of
individual U(1) charged particles $m\,Q^1$. 
Moreover, large baryonic Q-balls 
may be efficiently 
produced in the early universe \cite{ks,emcd98} within a scenario of a
collapsing unstable Affleck-Dine condensate \cite{ad}.
In the MSSM the role of
the global charge is played by baryon or lepton number carried by
squarks or sleptons respectively \cite{drt}. 
%Moreover, large baryonic Q-balls 
%may be efficiently 
%produced in the early universe \cite{ks, emcd98} within a scenario of a
%collapsing unstable Affleck-Dine condensate \cite{ad}. 
Whereas L-balls
(carrying leptonic charge) are not expected to survive until the
present epoch for $Q \simle 10^{36}(M/{\rm TeV})^{4/5}$ due 
to emission of massless
neutrinos, B-balls (containing baryonic charge) with $Q \simge
10^{12}\,(M/{\rm TeV})^4$ are stable because of the largeness of
the nucleon mass. 
It has thus been proposed that B-balls produced
in the early universe are not only an attractive dark matter candidate
but may also be responsible for baryogenesis via partial evaporation of
B-balls in the early universe~\cite{ls}. (This is distinct from a
scenario
of evaporation of unstable B-balls~\cite{emcd99} 
which could be responsible for 
baryogenesis and the creation of neutralino dark matter.) 

In this paper we reinvestigate the
evaporation of B-balls. We find the efficiency of this process to be 
limited by the transport
of baryon number away from the soliton, resulting in somewhat
different
conclusions than drawn in prior work~\cite{ls}. 
%This implies that successful 
%baryogenesis requires orders of magnitude
%smaller B-balls than previously thought, with $Q$'s which are
%likely in conflict with existing detector limits on a large fractional
%contribution of B-balls to the critical density. 
%Though it seems unlikely that stable B-balls are responsible for both,
%the dark matter and baryogenesis, they may be responsible for either. 
We also give some previously unmentioned cosmological limits on
the abundances of B-balls.

B-balls may release baryonic charge via evaporation of
squarks at temperatures $T>m_\chi$,  where $m_\chi$ is the
squark mass \cite{ls}. Assuming that B-balls constitute
dark matter at present (with fractional contribution 
to the critical density of $\Omega_Q$), there number density $n_Q$ in the early
universe at temperature $T$ has been
\begin{equation}
n_Q \approx 0.8\times
10^{-9}\,g_\ast\frac{m_p}{M}\frac{\Omega_{Q}}{Q^p}\,T^3\;. 
\label{qdensity}
\end{equation}
Here $m_p$ is proton mass, $g_\ast$ is the number of relativistic
degrees of freedom at the considered epoch ($g_\ast = 3.909$ today), 
and we have assumed a
Hubble constant of $H = 70\,$ km s$^{-1}$ Mpc$^{-1}$.
Consider B-balls with the general properties for their mass
\begin{equation}
M_Q = \alpha MQ^p 
\label{massprop}
\end{equation}
and their radius
\begin{equation}
R_Q = \beta M^{-1}Q^{p/3}\;.
\label{radiusprop}
\end{equation}
For an essentially flat
effective potential (e.g. $U\to M^4$)
for $\Phi \gg m_\chi$ 
the coefficient $p=3/4$ \cite{dks} and the
numerical factors may be computed for large B-ball charge:
$\alpha\approx(4\pi/3)\sqrt{2}$, $\beta\approx 1/\sqrt{2}$
\cite{ls}. If, on the other side, the expression $U(\Phi)/\Phi^2$ has a
minimum at finite $\Phi$ then the B-ball mass scales with $p=1$ \cite{col}.

To trace the baryon evaporation rate of a population of B-balls
at high temperatures, we consider
the thermodynamical properties of a volume element containing only
one B-ball.
Let $Q$ be the charge of the B-ball surrounded 
by $N_B$ baryon number carrying particles in the diffuse plasma in a volume
$V$, such that the total charge in the volume element is $N_{\sm{tot}}=N_B+Q$. 
Then the Helmholtz free energy of the system  reads as follows
\begin{equation}
F = -VT^4\frac{7\pi^2g_B}{360} + \frac{3N_B^2}{g_B\,VT^2} 
+ \alpha MQ^p\; ,
\label{fenergy}
\end{equation}
where $g_B$ is the number of baryon charge carrying 
degrees of freedom 
%(e.g. $g_B=2$ for spin $1/2$  particles) 
and where we assumed a vanishing entropy, $S_Q\approx
0$, for the B-ball. Expression (\ref{fenergy}) further assumes that
baryon number in the plasma is carried by fermionic
relativistic degrees of freedom
(cf. \cite{fma}).
For fixed $N_{\sm{tot}}$, $V$, and $T$ 
such a configuration has to minimize the Helmholtz free energy
Eq. (\ref{fenergy}) in thermal and chemical equilibrium. 
This extremum may be obtained by varying the fraction of $N_{\sm{tot}}$ 
residing in the B-ball. One thus finds that only for charge density in
excess of  
\begin{equation}
{N_{\sm{tot}}\over V} \gg \left(\frac{g_B}{6}\alpha
MT^2\right)^{1/(2-p)} V^{(p-1)/(2-p)}\; 
\label{qcond}
\end{equation}
the existence of a B-ball is thermodynamically favorable.
In this case the B-ball carries
almost the entire charge of the volume element ($Q\sim N_{\sm{tot}}$).
Nevertheless, in chemical equilibrium a small fraction of baryon
number also resides in the plasma, with baryonic density 
\begin{equation}
n^{\sm{eq}}_B = \frac{g_B}{6}p\alpha MT^2Q^{p-1}\,.
\label{equidens}
\end{equation}
We stress that by constraining our considerations to the
thermodynamics of a finite domain with volume $V$ and baryon number 
$N_{\sm{tot}}$ we have not obtained the absolute minimum for the free
energy. Considering a larger domain, but with $N_{\sm{tot}}/V$ kept
fixed, one may always find a state of lower Helmholtz free energy. This is
also exemplified by the odd dependence of Eq. (\ref{qcond}) on volume (except
for $p=1$).
As already noted in \cite{ls}, in the extreme limit the 
lowest free energy state is reached when all the
baryon number of the universe is contained in one large B-ball. 

Nevertheless, considerations of partial chemical equilibrium are
appropriate for the initial conditions envisioned resulting from the
breakup of an Affleck-Dine condensate. Here B-balls of typical initial
charge $Q_0$ form, with negligible baryon number in the plasma~\cite{kk}. 
During the subsequent evolution of the universe the coalescence of
B-balls is not possible, such that a state of even lower
Helmholtz
free energy is not attainable, and only partial chemical equilibrium
between an individual B-ball and the plasma around it may be attained.
Following similar arguments Laine \& Shaposhnikov~\cite{ls} estimated
the baryon evaporation rate 
of B-balls via squark emission at high $T$ by
\begin{eqnarray}
\Gamma_{\sm{evap}} \equiv \frac{dQ}{dt} & = &
- \kappa(\mu_Q-\mu_{\sm{plasma}})T^2\,4\pi\,R_Q^2 \nonumber \\ 
& \approx & -\kappa^{\prime}\,4\pi\,R_Q^2\,n^{\sm{eq}}_B\quad {\rm for} \,\,\,\,
\mu_{\sm{plasma}} << \mu_Q\, ,
\label{evaprate}
\end{eqnarray}
where $\mu_Q$ is the chemical potential of the B-ball.
The second line of Eq.~(\ref{evaprate}) is operative
under the assumption
that the evaporated particles can be quickly transported away from the
B-ball surface in order to sustain a jump between chemical potentials
of the B-ball and the plasma
($\mu_Q-\mu_{\sm{plasma}}\neq 0$). Note that $\mu_Q =
\mu_{\sm{plasma}} = (p/6)\alpha M Q^{p-1}$ in chemical
equilibrium. The
constants $\kappa$ and $\kappa^{\prime}$ in Eq. (\ref{evaprate}) are
$\simle 1$, where for $\kappa^{\prime}\approx 1$ the evaporation rate
is at its physical upper limit, implying there is no dynamical
suppression of the evaporation and accretion of squarks, i.e. every
squark approaching a B-ball will be absorbed.
This is in contrast to the evaporation of quarks from B-balls which
is suppressed~\cite{ccgm} and thus of negligible importance. 
Nevertheless, baryon transport away from the B-ball is not as
efficient as envisioned in Ref.~\cite{ls}.
Baryons released from the B-ball surface, will establish chemical
equilibrium with the B-ball, if they are not able to escape the surface
layer and evaporation ceases. 

The transport mechanism of baryon number is by diffusion of
squarks and quarks in the hot
plasma. Solving the spherical diffusion equation with
diffusion constant $D$
\begin{equation}
\frac{\partial n_B(r,t)}{\partial t} = D\frac{1}{r}
	\left(\frac{\partial^2}{\partial r^2}r\,n_B(r,t)\right)\;,
\label{diffeqn}
\end{equation}
on the condition that the number density at the surface boundary
does not change with time ($n_B(R_Q,t) = n_B^{\sm{eq}}$), yields a
steady-state solution for the density $n_b$. We confirmed this result
numerically for all
radii $r \simle L_{\sm{d}}$, where $L_{\sm{d}}\approx\sqrt{D\,t}$ is the
diffusion length at time $t$. Therefore the particle flux through the B-ball
surface is constant and given by
\begin{equation}
\Gamma_{\sm{diff}} \equiv \frac{dQ}{dt} = -4\pi\,k\, R_Q\,D\,n^{\sm{eq}}_B\;
\label{diffrate}
\end{equation}
where the diffusion constant of relativistic squarks and quarks
in a hot plasma is $D \approx
a\,T^{-1}$ with $a\approx 6$ for quarks and 
$a\approx 4$ for squarks,
respectively \cite{jpt,dw}. We have determined the numerical constant
$k$ to be very close to unity, such that we will drop it in what follows.
Apart from the numerical results
the expression (\ref{diffrate}) can be 
motivated by assuming a constant flux $dQ/dt=4\pi\,R_Q^2\,\partial
n/\partial r$ and approximating $\partial n/\partial r$
by $n^{\sm{eq}}_B/\Delta r$. The only time independent lengthscale in
this system is the B-ball radius $R_Q$, so $n_B^{\sm{eq}}/\Delta
r=n_B^{\sm{eq}}/R_Q$. 

By comparing the rates (\ref{evaprate}) and (\ref{diffrate})
\begin{equation}
\Gamma_{\sm{diff}}/\Gamma_{\sm{evap}}
 = \frac{D}{\kappa^{\prime}\,R_Q} \sim \frac{M}{T}Q^{-p/3}
\end{equation}
it is obvious that, for large B-balls, the diffusive transport is
orders of magnitude less
efficient than the evaporation of baryons from the B-ball. Since the 
evaporated
baryons are still within the surface layer of the B-ball, the
B-ball is at close to chemical equilibrium with the surrounding plasma and the
charge emission rate (\ref{evaprate}) must be replaced by
(\ref{diffrate}). 

Evaporation of squarks from B-balls is only efficient for
temperatures above the squark mass $m_\chi\approx
0.1 - 1\;\mbox{TeV}$. For temperatures below this mass 
the evaporation rate is exponentially suppressed by the Boltzmann
factor $\exp(-m_\chi/T)$. By integrating Eq. (\ref{diffrate}) one may
calculate the number of emitted baryons from a single B-ball until 
evaporation becomes inefficient at temperature $T_{\sm{fin}} \approx m_\chi$ 
\begin{equation}
\Delta Q 
%%= Q_0\left(1-\left(1-Q_0^{-(2-\frac{4}{3}p)}
%%\,b\,\frac{M_0}{T_{\sm{fin}}}\right)^{\frac{1}{\frac{4}{3}p-1}}\right)
\approx b\,\frac{M_0}{T_{\sm{fin}}}Q_0^{\frac{4}{3}p-1}\, .
\label{deltaq}
\end{equation}
Here $Q_0$ is the initial B-ball charge, and $M_0 =
(90/32 \pi^3 g_{\ast})^{1/2}M_{\sm{Pl}} \approx 3.7\times10^{18}/
\sqrt{g_{\ast}(T_{\sm{fin}})}\;\mbox{GeV}$ is given by the 
time-temperature relation $t = M_0\,T^{-2}$ during a radiation
dominated universe with $g_\ast(T_{\sm{fin}})\approx 200$. 
Note that for the
interesting case $p=3/4$ the numerical constant in Eq. (\ref{deltaq})
$b = (4\pi/3)\,g_B \beta a p \alpha \approx 4.7\times 10^3$,
assuming $g_B\approx 72$, and $\Delta Q$ is independent of
the initial charge of the B-ball. To ensure that an initially formed
B-ball survives evaporation until present ($\Delta Q/Q_0\ll 1$), such a
B-ball must have an initial charge of $Q_0> 10^{18} - 10^{19}$.

Within a B-ball baryogenesis scenario the
number density of baryonic matter $n_B$ and the number density of
B-balls are related by
\begin{equation}
n_B = n_Q\,\Delta Q\;.
\label{bdensity}
\end{equation}
Combining Eq. (\ref{deltaq}), (\ref{bdensity}) and Eq. (\ref{qdensity})
one may calculate the baryon-to-photon ratio at the present epoch
\begin{equation}
\eta = \frac{n_B}{n_{\gamma}}\approx
	1.3\times 10^{-8}b\,\frac{m_p}{M}
\frac{\Omega_{Q}}{Q_0^{1-\frac{1}{3}p}}
	\frac{M_0}{T_{\sm{fin}}}\; .
\label{eta}
\end{equation}
For a B-ball with $p=3/4$, $M \sim 1-10\;\mbox{TeV}$, and 
$T_{\sm{fin}}\sim 0.1 - 1\,$TeV it is
necessary to have $Q_0 \simge 9\times 10^{20} - 4\times 10^{23}$ to
obtain a baryon-to-photon ratio of $\eta \approx 3\times 10^{-10}$.
This should be compared to the range $Q_0 \sim 10^{22} -
10^{28}$ quoted in Ref. \cite{ls}. (Note that the above estimate has also
very different dependence
on the parameters $M$ and $T_{\sm{fin}}$ than the estimate by Ref. \cite{ls}.)
%(Note that if we would have made
%the same approximations regarding \lq\lq order untiy\rq\rq\
%constants as in Ref. \cite{ls}, but nevertheless would have taken into
%account baryon diffusion, we would have obtained
%$Q_0 \sim 10^{16} - 10^{18}$.) 
It is intriguing that the range of B-ball
charges which may yield successful baryogenesis is very close to the
observational lower limits $Q\simge 10^{21} - 3\times 10^{22}$ on the
charges of a significant B-ball galactic halo dark matter
population~\cite{kus}. 

A few comments concerning very large Q-Balls are of relevance. 
There are no detector limits on large B- and L-balls, either since their
current flux is extremely small, for stable solitons, or since they did
not survive to the present epoch, in the unstable case. There are,
however, some constraints on the existence of large Q-Balls in the
early universe. Consider first unstable $(p=1)$ B-balls. During their
decay they produce baryon inhomogeneities which may lead to a
scenario of inhomogeneous nucleosynthesis. It is well 
known~\cite{ahs,jf}, that
baryon lumps with baryon number in excess of

\begin{equation}
N_b \simge 10^{35} ({\eta_l/ 10^{-4}})^{-1/2}\quad
(\eta_l\gg 10^{-10})\;   
\end{equation}
can not homogenize by neutron diffusion
before the epoch of weak freeze-out. 
Here $\eta_l$ is the baryon-to-photon ratio in the
baryon-rich
region, and the value of $10^{-4}$ is of particular relevance.
It is expected that immediately after the B-ball decay $\eta_l^i\gg
10^{-4}$
(where we assume that baryon number is in form of diffuse baryons).
Neutrino heat conduction will subsequently expand the baryonic lump to
an asymptotic $\eta_l\approx 10^{-4}$, independent of
$\eta_l^i$, unless the
baryonic B-ball charge is in excess of $N_b \simge 10^{44}\eta_l^i$~\cite{jf}. 
If a fraction
$f_b \simge 10^{-2}$ of all baryons resides in such baryon number
enhanced regions, overproduction of $^4$He during nucleosynthesis
results. For $N_b \simge 10^{44}\eta_l^i$ similar constraints from $^4$He 
overproduction apply, but here even more stringent constraints may be
derived
from a possible overproduction of heavy elements. It is interesting to note,
that for $f_b \simle 10^{-2}$ B-balls with $N_b\simge 10^{35}$
may yield to the
production
of a primordial metallicity, without violating observational
constraints on the light element abundances~\cite{jfmk}.

A more speculative constraint may apply for large stable B-balls.
If their charge is in excess of $Q\simge 10^{44}\Omega_Q^{4/3} (M/{\rm
TeV})^{-4/3}$ their mean separation ($n_Q^{-1/3}$) at the QCD
transition at temperature $T\approx 100\,$MeV is $\simge 1\,$m. In the
case of a first-order QCD phase transition they may act as seeds for the
nucleation of hadronic phase. Depending on the amount of supercooling
which quark-gluon plasma may sustain before spontaneous nucleation
is efficient, and thus on the (three) surface free energies between 
the participants, hadronic phase, quark-gluon phase, and B-balls,
hadronic phase bubbles may only form around B-balls. If this is the
case,
the mean separation of baryon number enhancing quark-gluon plasma
bubbles
towards the end of the transition is such that baryon number
inhomogeneities with $N_b \simge 10^{35}$ of individual lumps (the baryon
number
within $\sim 1\,$m at $\eta\approx 3\times 10^{-10}$)
is large enough to yield a scenario of inhomogeneous nucleosynthesis.
Except for very narrow ranges in parameter space such a scenario is
typically
in conflict with observationally determined light element abundances.

In summary, we have reinvestigated a proposed scenario of baryogenesis
by the partial evaporation of stable B-balls in the early universe.
Under the assumption that the B-balls are the dark matter at the
present epoch, we have found that a successful baryogenesis scenario
by B-ball evaporation at high temperature
requires B-balls with baryon number $Q\approx 10^{20} - 5\times
10^{23}$, which is close to the observational lower limit on the charges of
a significant ($\Omega_Q\approx 1$) galactic B-ball population. Thus,
if stable B-balls are responsible for baryogenesis, and they constitute
the dark matter, they could be detected in the immediate future. 
We have also given some limits on the existence of larger B-balls in
the early universe.

\end{document}